\documentstyle[12pt]{article}
\newcommand{\be}{\begin{equation}}
\newcommand{\ee}{\end{equation}}
\newcommand{\bear}{\begin{eqnarray}}
\newcommand{\ear}{\end{eqnarray}}
\begin{document}
\begin{flushright}
HD---THEP---97--4
\end{flushright}
\vspace{1cm}
\begin{center}
{\bf \LARGE A Second Look at Masses and Mixings}\\ \vskip.5cm
{\bf \LARGE of Quarks and Leptons}

\vspace{1.5cm}
Berthold Stech\\

\medskip
Institut f\"ur Theoretische Physik, Universit\"at Heidelberg\\ Philosophenweg 16, D-69120 Heidelberg\\

\vspace{1cm}
\begin{abstract}
The masses of quarks and leptons in units of $m_t$ are expressed in powers of $\sigma=\sqrt{\frac{m_c}{m_t}}$. The regularities found suggest specific mass matrices. They reproduce all masses and mixings in striking agreement with observation. A 90$^o$ relative phase angle between the off-diagonal
elements of up and down quark matrices leads to a maximum magnitude for CP-violation and, for the phase angle $\gamma$ of the unitary triangle, to $\gamma\simeq 90^o$.
\end{abstract}
\end{center}
\newpage

Since the discovery of the top quark the masses of all quarks are known approximately. It is convenient to compare them in the $\overline {MS}$ scheme at
the common scale of 1 GeV. Expressed
in GeV units one has \cite{1}
\be\label{1}
\begin{array}{ll}
m_u(1\ {\rm GeV})=(4.8\pm1.5)10^{-3}&m_d (1\ {\rm GeV})=(8.5\pm1.5)10^{-3}\\
m_c(1\ {\rm GeV})=1.30\pm0.15&m_s(1\ {\rm GeV})=0.170\pm0.025\\ m_t(1\ {\rm GeV})=390\pm30&m_b(1\ {\rm GeV}) =6.4\pm0.4.\end{array}\ee
This knowledge makes it possible to look for regularities of quark and lepton masses
which may shed light on these mysterious quantities and on the even more mysterious mixing parameters and the CP-violating phase. One can also reconsider earlier suggestions \cite{2}-\cite{5} about the texture of mass matrices and their connection with the observed CP-violation.

My first observation concerns the up-quark masses. The mass ratios shown in (1) are consistent with the rule $m_u/m_c=m_c/m_t$\ , i. e.
\bear\label{2}
m_t:m_c:m_u&=&1:\sigma^2:\sigma^4\nonumber\\ {\rm with}\quad \sigma&=&\sqrt{\frac{m_c}{m_t}} = 0.058 \pm0.004.\ear
In case there is a unique force responsable for the pattern (\ref{2}) and which determines also the remaining masses and mixings, it appears natural to use $\sigma$ as an expansion parameter. I take then the mass ratios (\ref{2}) to hold up to
$O(\sigma)$ corrections.

Expressed in powers of $\sigma$ the down-quark mass ratios are consistent with the relations
\be\label{3}
m_b:m_s:m_d=1:\frac{1}{2}\sigma:8\ \sigma^3\ee The connection with the up-quark masses then is \bear\label{4}
m_b&=&\alpha_{bt}\ \sigma\ m_t\nonumber\\ m_s&=&\alpha_{bt}\ \frac{1}{2}\ m_c=\alpha_{bt}\ \frac{1}{2}\ \sigma^2 m_t\nonumber\\
m_d&=&\alpha_{bt}\ 8\ m_u=\alpha_{bt}\ 8\ \sigma^4 m_t\nonumber\\ {\rm with}&& \alpha_{bt}(1\ {\rm GeV}) = \frac {m_b}{\sigma\ m_t} = 0.28\pm0.03.\ear
It is again suggestive to assume that the relations (\ref{3}), (\ref{4}) hold apart from $O(\sigma)$ corrections. Notably, in (\ref{4}) the expression
for $m_b$ contains $\sigma$ linearly while $m_s/m_c$ and $m _d/m_u$ are independent of it.

Let's then also look at the masses of the charged leptons. To a good accuracy one has
\be\label{5}
m_\tau:m_\mu:m_e=1:\sigma:\frac{3}{2} \sigma^3\ee In (\ref{5}) the same powers of $\sigma$ appear as for the down-quark masses. Expressed in terms of $m_b$ and $m_t$ the lepton
masses can be written
\newpage
\bear\label{6}
&&m_\tau=\alpha_{\tau b}\ m_b=\alpha_{\tau b}\alpha_{bt}\ \sigma\ m_t\nonumber\\
&&m_\mu=\alpha_{\tau b}\ \sigma\ m_b=\alpha_{\tau b} \alpha_{bt}\ \sigma^2\ m_t\nonumber\\
&&m_e=\alpha_{\tau b}\ \frac{3}{2}\ \sigma^3\ m_b =\alpha_{\tau b}\alpha_{bt}\ \frac{3}{2}\ \sigma^4\ m_t\nonumber\\ &&{\rm with}\
 \alpha_{\tau b}(1\ {\rm GeV})=\frac{m_{\tau}} {m_{b}} = 
0.28\pm0.02.\ear $\alpha_{bt}$
and $\alpha_{\tau b}$ are scale-dependent. $\alpha_{\tau b}$ approaches $\simeq1$
at unification energies \cite{a}. The numerical equality of $\alpha_{\tau b}$ with $\alpha_{bt}$ at 1 GeV (see (\ref{4}) and (\ref{6})) is notable but could be entirely fortuitious. 

The next task is now to construct mass matrices which appear reasonable from the point of view of the dominance of the top quark and the structures found so far. 

In the standard model as well as in more general models which have no flavour-changing right-handed currents the mass matrices can be taken to
be hermitian matrices without loss of generality \cite{5}, \cite{6}.
It is of advantage to take the hermitian up-quark mass matrix with eigenvalues corresponding to (\ref{2}) not in its diagonal form but close to it such that the 33 element is equal to $m_t$ up to corrections of order $\sigma$, and all other elements are of first or higher order in $\sigma$. The motivation is clear: off-diagonal matrix elements with a lower power of $\sigma$
than diagonal elements and eigenvalues may have generated these by a higher order or step by step process. To express mass matrices in terms of powers of small parameters has been suggested by several authors \cite{b}.
Accordingly, the up-quark mass matrix $M^U$ is expected to exhibit the pattern \be\label{7}
M^U=\left(\begin{array}{ccc}
\sim\sigma^4&\sim\sigma^3&\sim\sigma^2\\ \sim\sigma^3&\sim\sigma^2&\sim\sigma\\
\sim\sigma^2&\sim\sigma&1\end{array}
\right)m_t\ .\ee
If the down-quark mass matrix $M^D$ is indeed strongly influenced by $m_t$ and $\sigma$ as suggested by (\ref{3}), (\ref{4}), its 33 element to leading order will be
\be\label{8}
m^D_{33}=\alpha_{bt}\ \sigma\ m^U_{33}=\alpha_{bt}\ \sigma\ m_t\ee with all other elements being of higher order. Consequently, as a first result, all mixing angles -- apart from the CP-violating phase angle -- must be of order $\sigma$ or smaller. In fact, the experimentally determined numerical values \cite{7} for the Kobayashi-Maskawa matrix elements are consistent with the following expressions to be ``derived'' further below:
\be\label{9}
|V_{us}|\simeq 4\sigma,\quad |V_{cb}|\simeq\frac{\sigma} {\sqrt2},
\quad|V_{ub}|\simeq\sigma^2\ .\ee
Numerically, (\ref{9}) gives -- taking $\sigma=0.056$ -- \be\label{10}
|V_{us}|\simeq0.22,\quad|V_{cb}|\simeq0.040,\quad|V_{ub}|\simeq0.0031 \ee
in good agreement with present knowledge. In view of these relations it is appropriate to take $\sigma$ instead of
the Cabibbo angle as the expansion
parameter in a Wolfenstein-type parametrization of the Kobayashi-Maskawa matrix which then reads \be\label{11}
KM=\left(\begin{array}{ccc}
1-8A^2\sigma^2&4A\sigma&Be^{-i\delta}\sigma^2\\ -4A\sigma&1-(8A^2+\frac{C^2}{4})\sigma^2&\frac{C}{\sqrt2} \sigma\\
(4\frac{AC}{\sqrt 2}-Be^{i\delta})\sigma^2& -\frac{C}{\sqrt2}\sigma& 1-\frac{C^2\sigma^2}{4}\end{array} \right)\ee
with $A\simeq1,\ B\simeq 1,\ C\simeq1$.
The new parametrization in terms of $\sigma$ has the advantage to govern besides the KM-matrix also the quark and charged lepton mass ratios.

Let us now discuss a possible
form of the hermitian up-quark mass matrix. For convenience we will take $m_{13}^U=m_{31}^{U*}$ and
$m_{23}^U=m_{32}^{U*}$ to be real and positive numbers as can always be achieved by a proper choice of the quark phases. In view of the dominance of the top quark it appears natural to set 
\be\label{12}
m^U_{22}=\pm\frac{(m^U_{23})^2}{m_{33}},\quad m^U_{12}=\pm \frac{m^U_{13}\cdot m^U_{32}}{m_{33}},\quad m^U_{11}=\pm\left(\frac{m^U_{12}\cdot m^U_{21}}{m_{22}} +\frac{m^U_{13}\cdot m^U_{31}}{m_{33}}\right)\ .\ee $m_{22}^U$, $m_{12}^U$ and $m_{11}^U$ are then quadratic, cubic and quartic in $\sigma$, respectively, as required by (\ref{7}).

By fixing the eigenvalues of $M^U$ (apart from signs and to leading order in $\sigma$) according to (\ref{2})
the eqs. (\ref{12}) have the consequence \bear\label{13}
&&m^U_{11}=0,\quad m^U_{12}=\pm\frac{1}{\sqrt2}\sigma^3\ m_t,\quad 
m^U_{13}=
\sigma^2\ m_t\nonumber\\
&&m^U_{22}=
-\frac{1}{2}\sigma
^2\ m_t,\quad m^U_{23}=\frac{1}{\sqrt2}\ \sigma\ m_t,\quad , \quad m^U_{33}=m_t\ .\ear

Thus, the up-quark mass matrix is, in our phase convention, a real and symmetric matrix with a rather simple structure\footnote{$m^U _{12}$ is chosen to be positive. A negative value still allowed by (\ref{12}) does not give acceptable results for the KM matrix.}: \be\label{14}
M^U=\left(\begin{array}{ccc}
0&\frac{1}{\sqrt2}\sigma^3&\sigma^2\\
\frac{1}{\sqrt2}\sigma^3&-\frac{1}{2}\sigma^2&\frac{1}{\sqrt2}\sigma\\ \sigma^2&\frac{1}{\sqrt2}\sigma&1\end{array}\right)m_t\ \ .\ee The zero at the $m_{11}$ position is of particular interest. It was often postulated following the classical paper by Weinberg \cite{2}. Here it is a consequence of (\ref{12}) and the negative sign of $m_{22}^U$.

The down-quark mass matrix $M^D$
should not have off-diagonal elements of first order in $\sigma$ as we remarked earlier. Therefore, $m_{22}^D$ agrees to leading order in $\sigma$ with the corresponding eigenvalue: According to (\ref{4}) and taking the sign of the eigenvalues of (\ref{14}) into account, one has\footnote{In ref. \cite{5} a common proportionality factor $\alpha$ for the ratio of the diagonal elements of down and up quark mass matrices was postulated. Now, there is a modification for the 3rd family. As seen from (\ref{8}) the ratio $m_{33}^D/m^U_{33}$ contains the additional factor $\sigma$.} \be\label{15}
m_{22}^D=-\frac{\alpha_{bt}}{2}\sigma^2m_t=\alpha _{bt}\cdot m_{22}^U\ .\ee

With $m^D_{33}$ of order $\sigma$ and $m^D_{22}$ given by (\ref{15}) the simplest choice for the down-quark mass matrix is obtained by decoupling the $b$-quark from $s$- and $d$-quarks, i.e. by setting $m^D_{13}=m^D_{23}=0$. The down-quark mass matrix then remains invariant under an arbitrary change of the $b$-quark phase. Taking also $m^D_{11}=0$ to have the lightest particle mass generated by off-diagonal elements as in the case of $m_u$ the down-quark mass matrix is given by \be\label{16}
M^D=\left(\begin{array}{ccc}
0&s_D\sigma^3&0\\
s_D^*\sigma^3&-\sigma^2/2&0\\
0&0&\sigma\end{array}\right)
\alpha_{bt}\ m_t\ee
with $|s_D|=2$.
The value $|s_D|=2$ gives -- to leading order in $\sigma$ and apart from a removable negative
sign -- the desired down-quark masses (\ref{3}), (\ref{4}). 

For the charged lepton matrix $M^E$
we use the same texture as for $M^D$ as suggested by models of grand unification. Then $M^E$ is given by \be\label{17}
M^E=\left(\begin{array}{ccc}
0&s_E\sigma^3&0\\
s^*_E\sigma^3&-\sigma^2&0\\
0&0&\sigma\end{array}\right)
\alpha_{\tau b}\alpha_{bt}\ m_t\ee
with $|s_E|=\sqrt{\frac{3}{2}}$.

To study the CP properties resulting from (\ref{14}), (\ref{16}) we follow
C. Jarlskog \cite{6} and consider the
commutator of the up and the down quark mass matrices and its determinant. To lowest orders in $\sigma$ the commutator reads
\be\label{18}
[M^U,M^D]=\left(\begin{array}{ccc}
\frac{s_D^*-s_D}{\sqrt2}\sigma^6
&\frac{1}{2}(s_D-\frac{1}{\sqrt2})\sigma^5&\sigma^3\\ -\frac{1}{2}(s_D^*-\frac{1}{\sqrt2})
\sigma^5&-\frac{s^*-s}{\sqrt2}\sigma^6&\frac{1}{\sqrt2}\sigma^2 \\
-\sigma^3&-\frac{1}{\sqrt2}\sigma^2&0\end{array}\right) \alpha_{bt}\ m^2_t\ee
and the determinant
\be\label{19}
Det\ [M^U,M^D]=\frac{1}{\sqrt2}(s_D^*-s_D) \sigma^{10}\alpha^3 _{bt}m_t^6\ .\ee
The determinant of the commutator measures the magnitude of CP-violation.

Because $|s_D|$ is fixed, it is immediately seen that the maximum possible amount of CP-violation will occur when the relative phase angle between the off-diagonal elements of the up and down quark mass matrix -- in our basis the phase of $s_D$ -- will be $\pm\pi/2$:
\be\label{20}
s_D/|s_D|=\pm i\ .\ee
As in ref. \cite{5} and as long as experiments do not disprove it, I will stick to this attractive possibility.\footnote{In ref. \cite{5} it was suggested that there exists a current quark basis in which all off-diagonal elements of the down-quark mass matrix are purely imaginary (using a diagonal up-quark mass matrix). To lowest order in $\sigma$ the requirement (\ref{20}) is fully equivalent to this earlier proposal, as can be seen by transforming $M^U$ and $M^D$ such that $M^U$ is diagonal and by using appropriate quark phases.} 

The mass matrices (\ref{14}), (\ref{16}), (\ref{17}) depend -- apart from the overall scale factor $m_t$ -- on three parameters only: $\sigma=\sqrt{\frac{m_c}{m_t}}$, $\alpha_{bt}=
\frac{m_b}{\sigma m_t}$ and ${\alpha}_{\tau b}=\frac{m_{\tau}}{m_b}$ with $ \alpha_{bt} \simeq\alpha_{\tau b}$ at 1\ {\rm GeV}.
But, of course, one has to keep in mind that their construction was based on mass ratios abstracted from (\ref{1}).
Using in addition the condition for maximal CP-violation (\ref{20}), the diagonalization of the mass matrices gives 6 quark masses, 3 charged lepton masses and the 4 quark-mixing angles. To lowest order in $\sigma$
one obtains, beside the mass relations we started from, the formulae already exhibited in (\ref{9}) and for the phase angle $\gamma$ of the triangle diagram $\gamma=\pi/2$ (by taking the minus sign in (\ref{20})).

Let us forget now the ``derivation'' of the mass matrices and simply propose them in the precise 
form given by (\ref{14}), (\ref{16}) (\ref{17}). The special factors multiplying the powers of $\sigma$ as well as the zero's in the mass matrices are now supposed to be valid at some high energy scale. At low energies modifications occur depending on the scale difference and on the group structure between the two scales. However, apart from the scale dependence of $m_t$, $\alpha_{bt}$ and $\alpha_{\tau b}$, these modifications are generally very mild. We will ignore them here and calculate now the masses and mixings to all orders of $\sigma$ by diagonalizing the mass matrices (14), (16), (17) numerically i.e. without expanding in powers of $\sigma$. Not using a best fit, but just taking the central values from (\ref{1}) for $m_t$, $\alpha_{bt}$, $\alpha_{\tau b}$ and $\sigma$, namely, \be\label{21}
m^U_{33}=390\ ,\quad \alpha_{bt}=\alpha_{\tau b}=0.28\ ,\quad \sigma=0.058\ee
one obtains the masses (in GeV units)
\be\label{22}
\begin{array}{lll}
m_t(1\ {\rm GeV}) = 391 & m_c(1\ {\rm GeV}) = 1.31 & m_u (1\ {\rm GeV}) = 4.4\cdot 10^{-3} \\
m_b(1\ {\rm GeV}) = 6.33 & m_s(1\ {\rm GeV}) = 0.19  & m_d(1\ {\rm GeV}) = 9.4\cdot10^{-3}\\ 
m_\tau(1\ {\rm GeV}) = 1.77 & m_\mu(1\ {\rm GeV}) = 0.103 & m_e(1\ {\rm GeV}) = 0.52 \cdot10^{-3}\end{array}\ee
and for the Kobayashi-Maskawa matrix elements and the angles $\alpha$, $\beta$, $\gamma$ of the unitarity triangle\bear\label{23}
&&|V_{us}| = 0.216\ , \quad|V_{cb}| = 0.041\ ,\quad|V_{ub}| = 0.0034 \ .\quad|V_{td}| = 0.0094\ , \nonumber\\
&&\quad\alpha = 70^o, \quad\beta = 20^o, \quad\gamma = 90^o \ .\ear In conventional notation of the KM matrix $V_{ub}$ and $V_{td}$ read
\be\label{24}
V_{ub} = -0.0034 i\ ,\quad V_{td} = -0.0033 i+0.0088\ .\ee The agreement with known data \cite{7} is quite encouraging. 

I can say little about lepton (neutrino) mixing. It depends of course on the neutrino mass matrix for which we have no good information at present. With the charged lepton matrix in the form (\ref{16}) the neutrino mass matrix depending on $m_t$ and $\sigma$ will certainly be close to a diagonal matrix. Thus, only small mixing angles, the largest definitely smaller than the Cabibbo angle, can be expected. The similarity of $M^E$ with $M^D$ suggests CP-violation and $s_E/|s_E|=\pm i$ if
maximal CP-violation occurs in the quark sector. 

In conclusion I can state: The masses and mixings of quarks and leptons appear to have a common origin which suggests mass matrices of a particularly simple form. The elements of the mass matrices are governed by powers of the small parameter $\sigma=\sqrt{\frac{m_c}{m_t}}$, which is close to $\frac{2m_s}{m_b}$ and $\frac{m_\mu}{
m_\tau}$. The factors in front of the powers of $\sigma$ remind of Clebsch-Gordan or normalization-type numbers. Their knowledge may possibly help to find the underlying dynamics. For instance, one might interpret these factors as manifestations of a family-type symmetry in a similar way as the numbers obtained from models studied in ref. \cite{c}. 

\bigskip
\noindent{\bf Acknowledgement:} It is a pleasure to thank D. Gromes, M. Jamin, O. Nachtmann, M.G. Schmidt and C. Wetterich for valuable
remarks.

\end{document}